\documentclass[useAMS,usenatbib,usegraphicx]{mn2e}
\usepackage{multirow}
\title[Predicted properties of Galactic and Magellanic Classical Cepheids in the SDSS filters]{Predicted properties of Galactic and Magellanic Classical Cepheids in the SDSS filters}
\author[]{Di Criscienzo, M.$^{1}$ , Marconi, M.$^{1}$,  Musella, I.$^{1}$,  Cignoni, M.$^{2}$, Ripepi, V.$^{1}$\\
$^{1}$ INAF, Osservatorio Astronomico di Capodimonte, Via Moiariello 16, 80131 Napoli, Italy.\\
$^{2}$ INAF, Osservatorio Astronomico di Bologna, Via Ranzani 1, 40127 Bologna, Italy\\
}
\begin{document}

\date{Accepted ??}

\maketitle

\label{firstpage}

\begin{abstract}
We present the first extensive and detailed theoretical scenario for the interpretation of Cepheid properties observed in the SDSS filters. Three sets of nonlinear convective pulsation models, corresponding to the chemical compositions of Cepheids in the Milky Way, the Large Magellanic Cloud and the Small Magellanic Cloud respectively, are transformed into the SDSS bands by relying on updated model atmospheres.
The resulting observables, namely the instability strip boundaries and the light curves, as well as the Period-Luminosity, the Wesenheit and the Period-Luminosity-Colour relations, are discussed as a function of the metal content, for both the fundamental and the first overtone mode.
The fundamental PL relations are found to deviate from linear relations when computed over the whole observed Cepheid period range, especially at the shorter wavelenghts, confirming previous findings in the Johnson-Cousins bands.
The obtained slopes are found to be mildly steeper than the ones of the semiempirical and the empirical relations available in the literature and covering roughly the same period range, with the discrepancy ranging from $\sim$ 13 $\%$ in $u$ to $\sim$ 3 $\%$ in $z$.

 \end{abstract}


\section{Introduction}
Thanks to their characteristic Period-Luminosity (PL) and
Period-Luminosity-Color (PLC) relations, Classical Cepheids are the
most important stellar standard candles to estimate accurate distances
within the Local Group, reaching  $\sim$ 30 Mpc  from space
observations. These Cepheid based distances provide an absolute
calibration of the extragalactic distance scale and two HST projects
have been devoted to this issue, obtaining an estimate
of the Hubble constant  \citep[see e.g.][]{f01,s01}.
Moreover, Cepheid properties are very important tools to investigate
the physics and the evolutionary behaviour of intermediate mass
stars \citep[see e.g.][and references therein]{m09}. Even if the Cepheid PL and PLC relations are  traditionally studied in
the standard Johnson-Cousin BVRIJK magnitude system,  the broadband filters employed in the Sloan
Digital Sky Survey \citep[SDSS; see e.g.][]{ab03} are becoming more
and more
popular in the context of current and future projects.
 For example, observations with the VLT Survey
 Telescope (VST),  the Canada-France-Hawaii Telescope (CFHT), the
 Pan-STARRS1 survey (Kaiser 2004), the Dark Energy Survey2 (DES; Abbott et al. 2006), and the Large Synoptic Survey Telescope3 (LSST; Tyson 2002) are being  done in the Sloan filters and from programs including a time-domain component,  a large number of variable stars, including Cepheids, is expected. 
In particular, we are deeply involved in the INAF-GTO approved VST survey STEP \citep[SMC in Time-Evolution of a Prototype interacting late-type dwarf galaxy; PI: V. Ripepi,][]{ripepiSTEP}). This project will  include $g_{SDSS}$ and $i_{SDSS}$ time-series photometry of the Bridge, connecting the two Magellanic Clouds, in order to study its stellar populations  using the
variable stars as tracers. Thanks to approved and submitted ESO proposal we plan to obtain 20-25 phase points in $g_{SDSS}$  and $i_{SDSS}$ on 32 squared degrees to identify and study variables including both Classical Cepheids and RR Lyrae.
Periods and mean $g_{SDSS}$ and $i_{SDSS}$ magnitudes will be obtained from these data and accurate PL and PLC relations will be derived.\\
In order to guarantee a reliable interpretation  of these and other Cepheid data and to assure a comparison between theory and observations,
in this paper we provide a theoretical scenario to interpret the properties of Classical Cepheids in the SDSS filters, for the characteristic chemical composition shown by these pulsators in the Milky Way and the Magellanic Clouds.\\
The organization of the paper is the following.
In Section \ref{puls} we present the nonlinear convective pulsation models
developed by our team, whereas in Section \ref{sdss} the adopted model
atmospheres to transform the theoretical quantities and the bolometric
light curves into the SDSS filters are discussed. The results are
shown in Section \ref{puls_sdss} and in Section \ref{comp} we present the comparison between model predictions and
previous results available in the literature. The
Conclusions close the paper.

\section{Pulsation models} \label{puls}
The adopted nonlinear convective pulsation models are taken from
\citet{bms99,b01,b02} for the chemical compositions representative of
Galactic and Magellanic Cepheids, namely $Z=$0.02, $Y=$0.28 (Milky Way),
$Z=$0.008, $Y=$0.25 (LMC) and $Z=$0.004, $Y=$0.25 (SMC).
For each selected stellar mass in the range from 3.5 $M_{\odot}$ to 11
$M_{\odot}$ a luminosity level based on an evolutionary Mass-Luminosity relations that neglects both core overshooting and mass loss \citep[see][and references therein]{b01}. Previous theoretical investigation has shown that the inclusion of mild overshooting (corresponding to an increase in the intrinsic luminosity of 0.25 dex, at fixed mass,  according to Chiosi, Wood and Capitanio 1993) implies a decrease of about 0.2 mag in the distance moduls inferred from theoretical Wesenheit and PLC relations (see e.g. Caputo, Marconi, Musella 2000).\\
The properties of adopted pulsation models are summarized in Table 1. For each combination of chemical composition, mass and luminosity,
a wide range of effective temperatures has been explored to define the
position of the instability strip boundaries for both the fundamental
and the first overtone mode.

\begin{table}
\begin{center}
\caption{Parameters of the computed pulsational models adopted in this paper.The last four columns are the temperatures in Kelvin of the first overtone
blue edge (FOBE), the fundamental blue edge (FBE), the first overtone
red edge (FORE) and the fundamental red edge (FRE) respectively.}
\begin{tabular}{l|c|c|cccc}
\hline
\hline
Z&M/M$_\odot$&LogL/L$_\odot$&FOBE&FBE&FORE&FRE\\
\hline
\multirow{14}{*}{0.004}
&3.25 & 2.490  &   6650 & 6025 &  5950 & 5825 \\
&3.50 & 2.610  &   6650 & 6025 &  5950 & 5775 \\ 
&3.80 & 2.740  &   6650 & 6025 &  5950 & 5675 \\ 
&4.00 & 2.820  &   6650 & 5950 &  5850 & 5625 \\ 
&5.00 & 3.070  &   6450 & 5950 &  5850 & 5550 \\ 
&5.80 & 3.319  &   6450 &  -   &  5950 &  -   \\ 
&5.50 & 3.400  &   6450 &  -   &  5950 &  -   \\ 
&7.00 & 3.650  &   6450 &  -   &  5950 &  -   \\ 
&7.00 & 3.650  &   6050 & 5950 &  5550 & 5150 \\ 
&7.15 & 3.730  &   5950 & 5950 &  5650 & 5050 \\ 
&7.30 & 3.760  &   5950 & 5950 &  5650 & 4950 \\ 
&7.45 & 3.795  &   5950 & 5950 &  5750 & 4950 \\ 
&9.00 & 4.000  &     -  & 5850 &   -   & 4850 \\ 
&11.0 & 4.400  &     -  & 5550 &   -   & 4350 \\ 
\hline
\multirow{14}{*}{0.008}
&3.25  &2.450  &  6750  &6125 & 6150 & 5875\\
&3.50  &2.570  &  6650  &5975 & 5950 & 5775\\
&3.80  &2.697  &  6650  &5950 & 5950 & 5675\\
&4.00  &2.777  &  6650  &5950 & 5850 & 5625\\
&5.00  &3.070  &  6400  &5950 & 5850 & 5450\\
&6.55  &3.550  &   -    &5950 &  -   & 5050 \\      
&6.70  &3.586  &   -    &5950 &  -   & 5050 \\  
&6.85  &3.620  &   -    &5850 &  -   & 5050 \\  
&7.00  &3.650  &  5850  &5850 & 5650 & 4950\\
&7.15  &3.688  &   -    &5850 &  -   & 4950 \\ 
&7.30  &3.720  &  5850  &5850 & 5750 & 4850\\
&7.45  &3.752  &   -    &5850 &  -   & 4850 \\
&9.00  &4.000  &   -    &5650 &  -   & 4650 \\
&11.0  &4.400  &   -    &5250 &  -   & 4150 \\
\hline
\multirow{14}{*}{0.02}
&  3.50 &  2.510&  6450 &  -  &  5950&  -   \\
&  4.00 &  2.720&  6550 &  -  &  5850&  -   \\
&  4.50 &  2.900&  6350 & 5850&  5850&  5450\\
&  5.00 &  3.070&  6150 & 5950&  5850&  5350\\
&  5.50 &  3.219&  5850 &  -  &  5750&    - \\
&  5.60 &  3.259&  5950 &  -  &  5850&   -  \\
&  6.25 &  3.420&   -   & 5550&  -   &  4950\\
&  6.50 &  3.480&   -   & 5650&  -   &  4950\\
&  6.75 &  3.540&   -   & 5575&  -   &  4850\\
&  7.00 &  3.650&   -   & 5450&  -   &  4650\\
&  9.00 &  4.000&   -   & 5150&  -   &  4350\\
& 11.00 &  4.400&   -   & 4850&  -   &  3850\\
\hline
\end{tabular}								
\end{center}				
\end{table}

Thanks to their nonlinearity these models allow us to predict the full
amplitude behaviour  and the variations of all the relevant quantities
along a pulsation cycle. Moreover the inclusion of a nonlocal
time-dependent treatment of convection \citep[see][for
details]{s82,bs94,bms99} is the key ingredient to predict the whole
topology of the 
instability strip\footnote{The existence of the red edge is related to the quenching
  effect of convection on pulsation.}, as well as the detailed
morphology of light, radius 
and radial velocity curves.
The physical and numerical assumptions adopted in the computation of
these models 
are discussed in detail in \citet{b98,bms99,bms00} and are not repeated here.
The modal stability has been investigated in the quoted papers for
both the fundamental and the first overtone modes, so that, moving from
higher to lower effective temperatures, the following four edges are
predicted for each selected chemical composition: the first overtone
blue edge (FOBE), the fundamental blue edge (FBE), the first overtone
red edge (FORE) and the fundamental red edge (FRE). Moreover,
theoretical atlas of bolometric light curves are available \citep[see
e.g.][]{bcm00} for each combination of chemical composition and mass (luminosity), varying the model effective temperature (period).

\section{Transformation into SDSS photometric system}\label{sdss}
In order to obtain the pulsation observables of the investigated Cepheid models into the SDSS bands, we have transformed both the
individual static luminosities and effective temperatures and the
predicted bolometric light curves into the corresponding filters,
using model ATLAS9 non-overshooting model atmospheres  \citep[][]{cast97a,cast97b},  convolved with the SDSS transmission functions.\\
In particular we used computed bolometric corrections and color
indices, as tabulated for different effective temperatures (3500 K$\le$
Teff $\le$ 50000K), surface gravities ($\log g$ from 0.0 to 0.5) and
metallicities ($[M/H]$=0.5, 0.2, 0.0, − 0.5, − 1.0, − 1.5, − 2.0
and −2.5, both for $[\alpha/Fe] =$ 0.0 and for $[\alpha/Fe] =$ 0.4 ),
available on the homepage of Fiorella Castelli (http://wwwuser.oat.ts.astro.it/castelli/colors/sloan.html).\\
For each selected chemical composition, interpolation through these tables allowed us to obtain the static magnitudes, as well as the light curves and the resulting magnitude-averaged and intensity-averaged mean magnitudes,  in the  $u_{SDSS}$, $g_{SDSS}$, $r_{SDSS}$, $i_{SDSS}$, and $z_{SDSS}$ filters ($u$,$g$,$r$,$i$,$z$ in the following  tables and figures)

\section{Predicted pulsation observables in the SDSS filters}\label{puls_sdss}
In this section we present the  results of the transformation of pulsation model predictions into the SDSS filters, providing the first theoretical scenario  to interpret observed Cepheid properties in these photometrical bands. 
\subsection{The light curves and the Hertzprung progression}
The bolometric light curves have been transformed into the $u$,
$g$, $r$, $i$, and $z$ filters. In
Figs. 1 we show some examples of the resulting
light curves in the $g$ filter\footnote{The light curves in the other filters and for other stellar masses are available upon request to the authors} for $Z=0.004$ ($M=7.15M_{\odot}$), $ Z=0.008$ ($M=7.15M_{\odot}$) and $Z=0.02$ ($M=6.5M_{\odot}$ ).
The behaviours shown in these plots confirm that the Hertzprung
progression \citep{H26,LW58}, that is the relationship linking
the phase of the bump and the pulsation period, occurs at $P \approx$ 11.5 d, 11.0 d and 9 d for models at $Z=$0.004, 008 and 0.02, respectively \citep[see e.g.][for an extensive
discussion of theoretical predictions for this phenomenon and the
comparison with observations]{bms00,mmf05}.

\begin{figure}
\includegraphics[width=9cm]{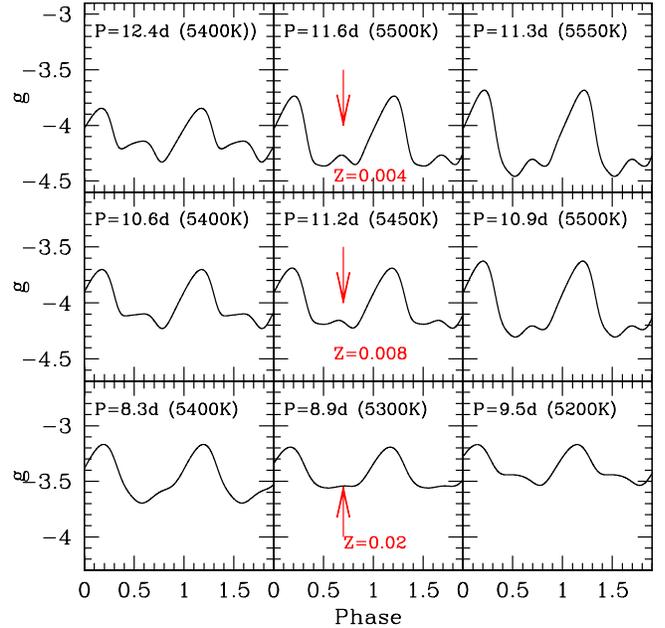} 
\caption{Theoretical atlas of light curves transformed in the
  $g$  band for  Z=0.004 and M=7.15$M_{\odot}$ (up), Z=0.008 and M=7.15$M_{\odot}$ (middle) and Z=0.02 and M=6.5$M_{\odot}$ (down). The arrow
  marks the position of the Hertzprung progression.}
\label{prog1}
\end{figure}

\subsection{The instability strip boundaries}
\begin{figure}
\includegraphics[width=9cm]{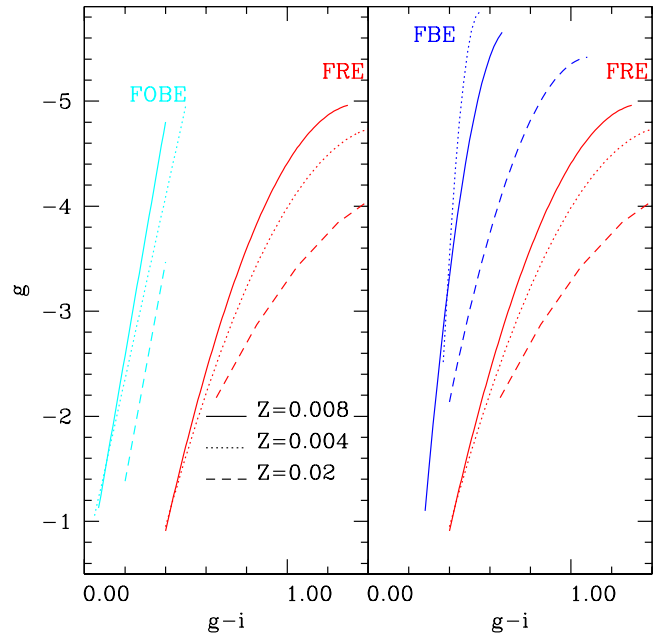} 
\caption{Predicted instability boundaries of the whole instability
  strip  (left panel) and for the fundamental mode only (right panel)
  in the intensity-averaged $g$ versus $g-i$ plane, for the  adopted
  model metal abundances.}
\label{figSTRIP}
\end{figure}

The  predicted instability strip boundaries reported in the last four
columns of Table 1 have been interpolated (we use a linear regression
for the FOBE and FORE and a quadratic relation for the FBE and FRE) 
and transformed into the SDSS filters. The results and shown in
Fig. \ref{figSTRIP} as a function of the adopted metallicity.

The left panel of this figure shows the behaviour of the extreme edges
(FOBE and FRE) of the pulsation region when moving from Z=0.004 to
Z=0.008 and Z=0.02. The right panel only refers to the fundamental
mode (FBE and FRE). \\
These plots indicate that also in the $g$ vs
$g$-$i$ plane the predicted instability strip gets
redder as the metallicity increases. In particular the nonlinear trend
and the metallicity effect, shown in
the right panel for the fundamental mode boundaries, are known to directly
reflect into the behaviour of the predicted fundamental PL relations
\citep[see][]{bccm99}.  In other words,  on the basis of the behaviours
noted in this figure, the PL relation in the $g$ filter is expected to
be nonlinear and metallicity dependent.

\subsection {Color-Color relations}

\begin{figure*}
\includegraphics[width=6cm]{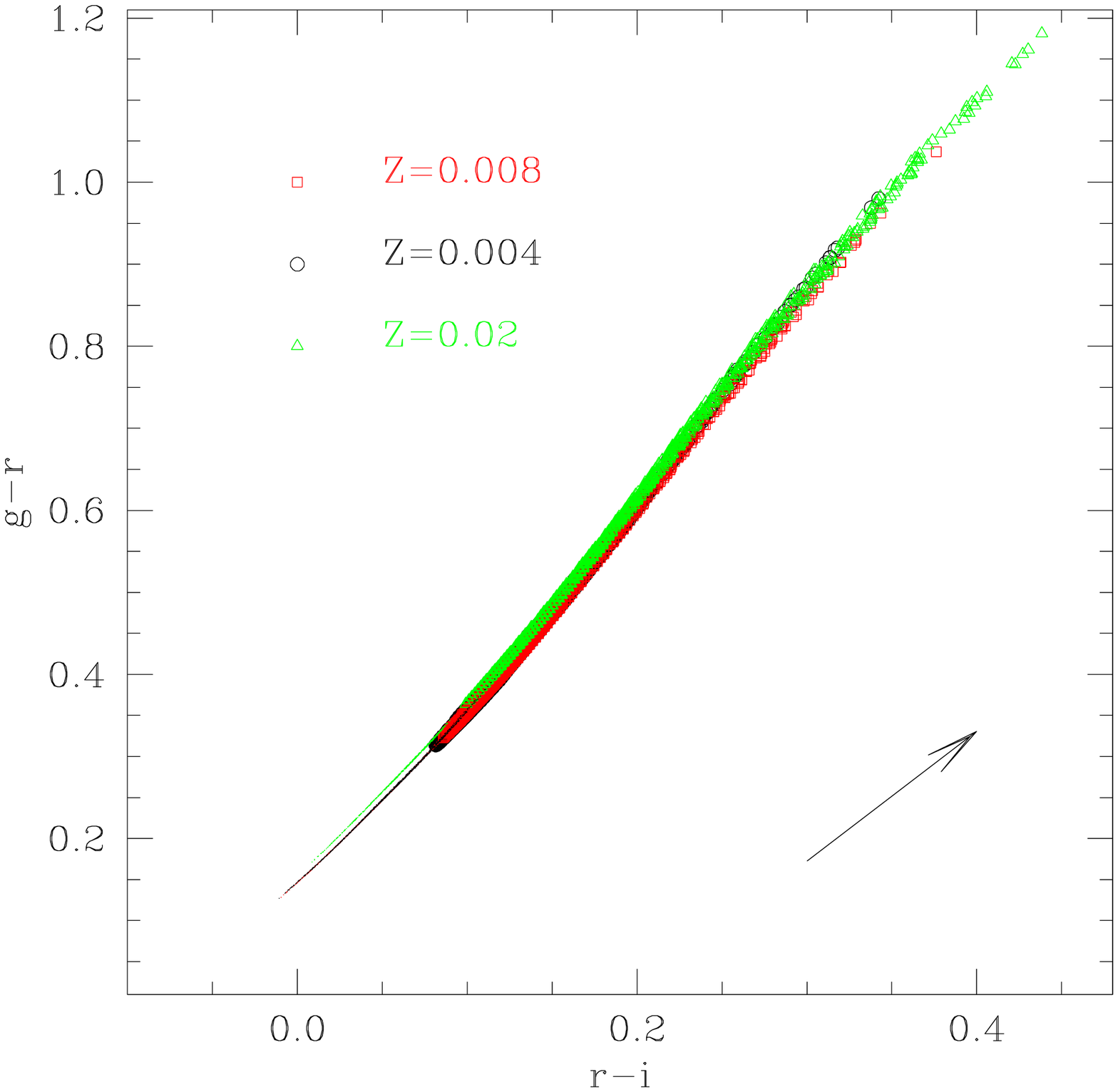}
\includegraphics[width=6cm]{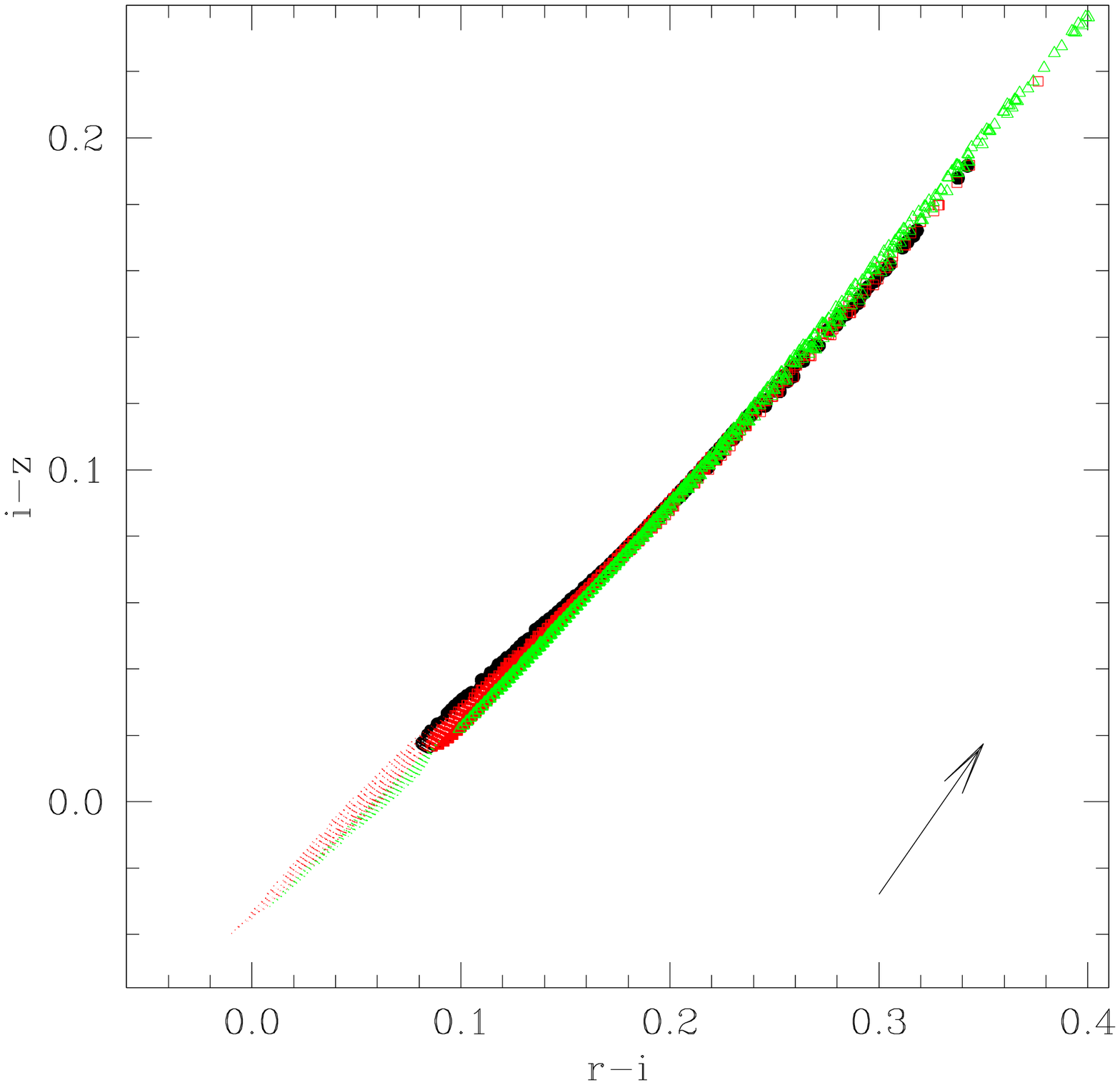}
\includegraphics[width=6cm]{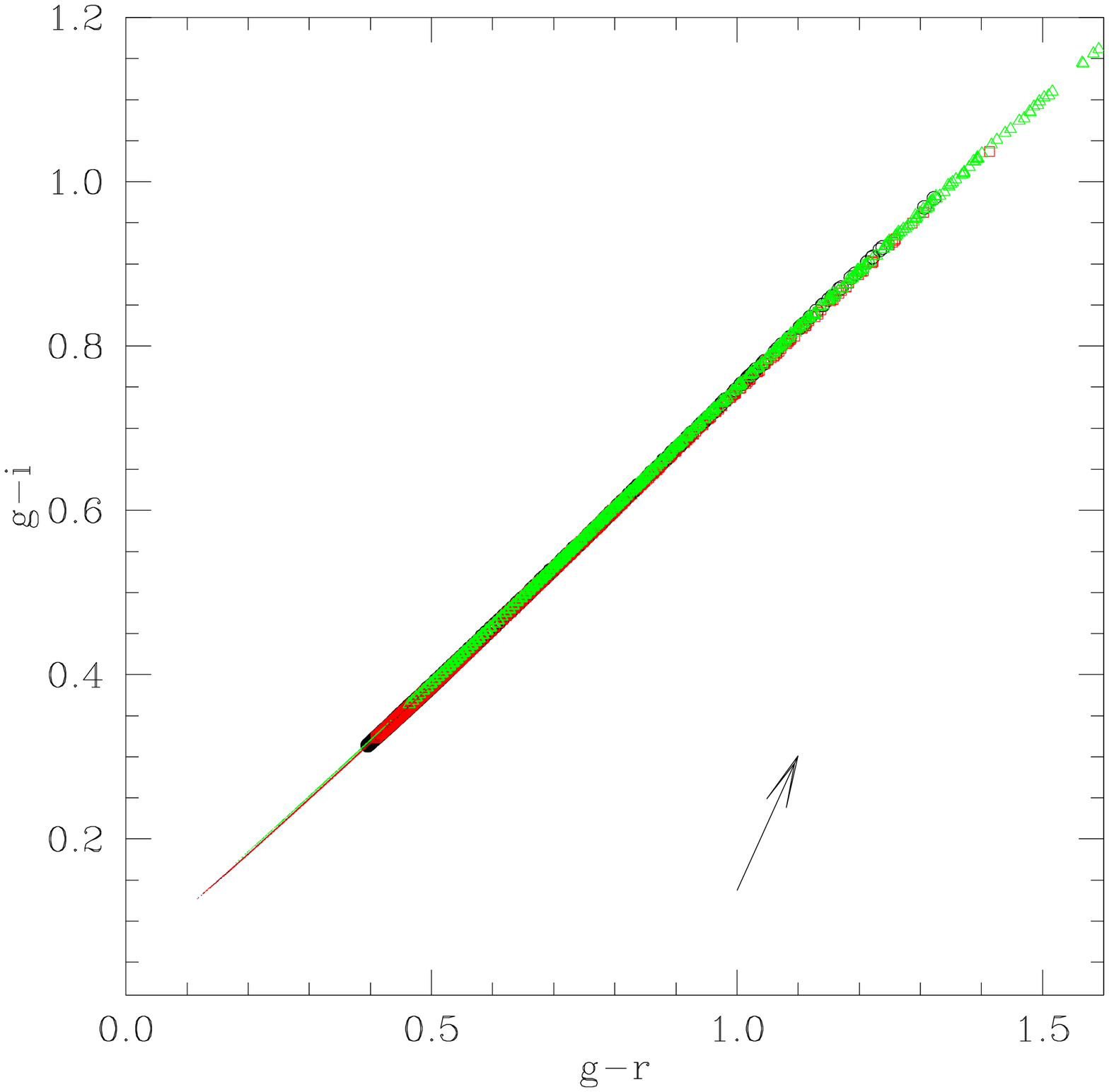}
\caption{Fundamental and first overtone (smaller points on the left side of each figure) models of different metallicities in colour - colour planes. A reddening vector for A$_V$ = 0.5 is also shown in each panel.}
\label{figCC}
\end{figure*}

The transformed multifilter light curves can be reported in different
colour – colour planes. In Figs. \ref{figCC} we show the theoretical fundamental (F) and First Overtone (FO) pulsators in the $r-i$ versus $g-r$, $i-z$ versus $r-i$ and $g-r$ versus $g-i$ diagrams for the adopted metallicities.\\
On the contrary of what happen in the $g-r$ versus $u-g$ plane, due to the significant sensitivity of the $u$ band on metal abundance, in  color-color plane reported in the figures  the relations are very narrow and the effect of metallicity is smaller. In particular, this effect becomes negligible for the  $g-i$ and $g-r$ combination of colors   suggesting that the comparison between theory and observations in this plane could be used to evaluate colour excesses.\\
The linear regression through the intensity-averaged mean magnitudes
of synthetic  F and FO pulsators (see subsection \ref{PL}) provides the following metal-dependent analytical colour – colour relations:\\
$$g-r = 0.12 + 2.34(r-i )  \ \ \ \    (\sigma =0.009) $$
$$i-z = -0.04 + 0.62(r-i) \ \ \ \  (\sigma= 0.002) $$
$$g-r = 0.04 + 0.70(g-i) \ \ \ \ (\sigma= 0.003)$$
for Z=0.004.\\
$$g-r = 0.12 + 2.35 (r-i ) \ \ \ \  (\sigma= 0.008)$$
$$i-z = -0.04 + 0.63 (r-i) \ \ \ \  (\sigma= 0.002)$$
$$g-r = 0.04 + 0.70(g-i) \ \ \ \  (\sigma= 0.002)$$
for Z=0.008.\\
$$g-r = 0.12 + 2.44 (r-i ) \ \ \ \ (\sigma= 0.008) $$
$$i-z = -0.04 + 0.67 (r-i) \ \ \ \  (\sigma = 0.003)$$
$$g-r = 0.04 + 0.71(g-i) \ \ \ \  (\sigma = 0.002)$$
for Z=0.02.\\

\subsection{Period - Luminosity relations }\label{PL}
\begin{figure}
\includegraphics[width=9cm]{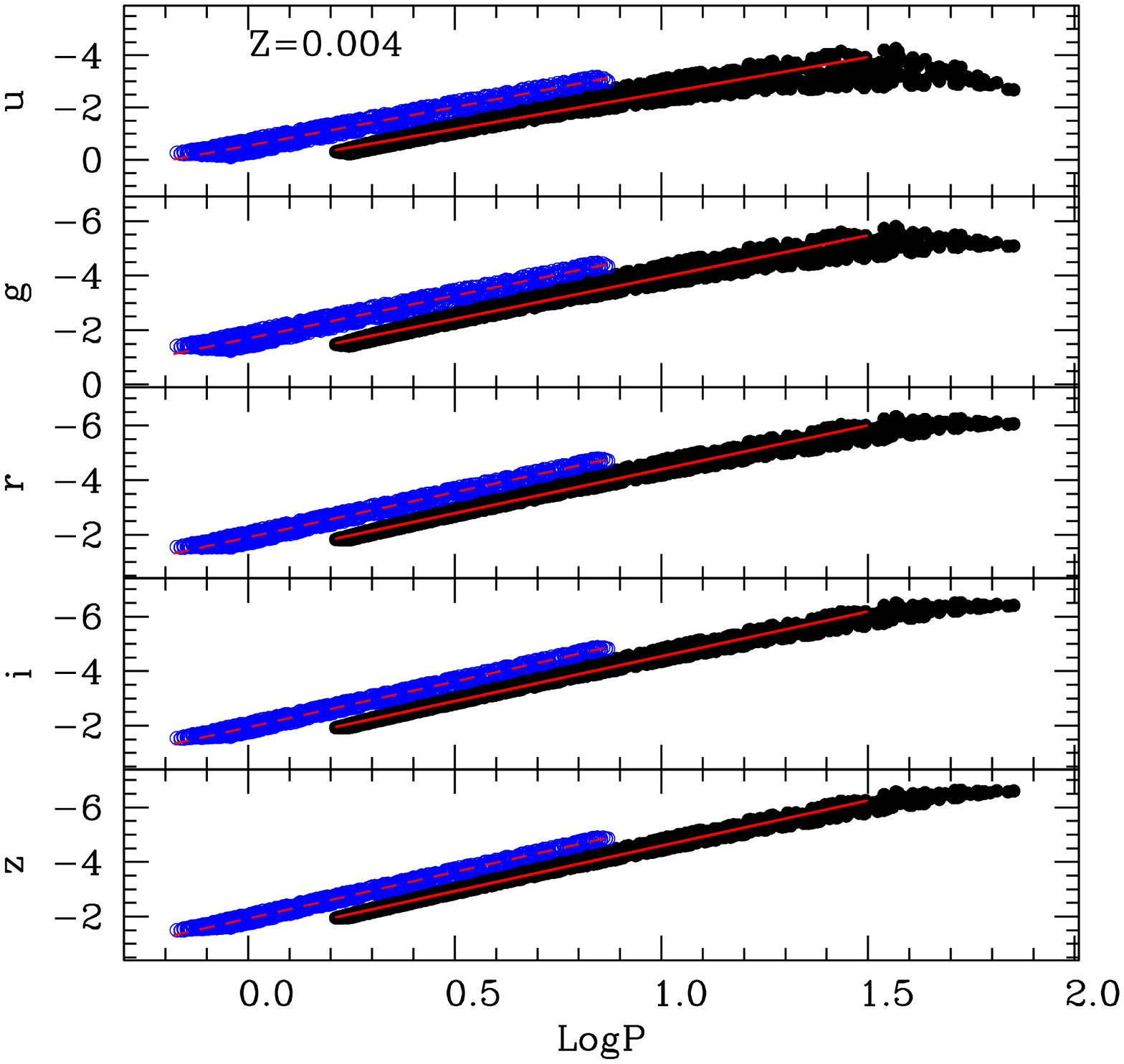} 
\caption{PL relations ($\log P \le$ 1.5, see Table 2) in different SDSS photometrical bands for fundamental (solid lines) and first overtone (dashed lines) models
with Z=0.004.The same figures for  Z=0.008 and Z=0.02, are given in the  on-line version of the paper.}
\label{figPLsmc}
\end{figure}
We used the synthetic distributions  of fundamental models computed by
\citet{cmm00} to derive the  PL relations for different photometric
bands listed in Table \ref{tabPL} for the three assumed chemical
compositions. We fitted with a linear regression M$_i$= a+b LogP the
predicted distribution of F pulsators for different assumptions on the
 period range, namely  $\log P \le$ 2.0, $\log P \le$ 1.5, $\log P \le$ 1.0 and $\log P\ge$ 1.0 to determine  the three corresponding slopes: $b_{all}$, $b_{\le 1.5}$, $b_{short}$ and $b_{long}$. 
The investigation of the difference between $b_{short}$ and $b_{long}$ is
relevant in the context of current studies in the literature
claiming the existence of a break at 10 days \citep[see e.g.][]{cmm00,mmf05,s04,n08}  
As shown in Fig. \ref{figPLsmc}  the PL relations
get steeper by increasing the filter wavelength, in  agreement with
the well-known empirical and theoretical results in Johnson-Cousins photometric bands \citep{mf1991,cmm00}, as well as with the semiempirical PL
relations in the SLOAN bands derived by \citep{nk07}.\\
Moreover, in agreement with the results by Bono (2010), the slopes of the
relations at LogP $\le$ 1.0 are steeper  than those for longer
periods, indicating a clear break of the PL relation at 10 days,
more evident for the Magellanic Cloud chemical compositions, in
agreement with empirical suggestions.
However, the amplitude of this effect decreases  as the filter
wavelength increases.
In the case of first overtone pulsators, that are intrinsically
limited to a shorter period range, linear relations over the
whole covered period range are derived.

The results reported in Table \ref{tabPL} clearly show the expected
flattening of the various fundamental PL relations when the
metallicity increases. Again, this trend decreases as the wavelength
increases.

The smaller, negligible passing from $Z=0.004$ to $Z=0.008$, sensitivity to metallicity of first overtone PL
relations is due to the much narrower instability strip that is also
hotter than the one of fundamental pulsators, and in turn less affected
by changes in the convective efficiency produced by metallicity
variations. 

\begin{table*}
\begin{center}
\caption{Predicted slopes of fundamental and first overtone PL
  relations in the SDSS filters. Columns from second to fourth report
  the slopes in three different assumptions on the period range, namely
  $\log P \le$ 2.0, $\log P \le$ 1.0 and $\log P\ge$
1.0 and   $\log P \le$ 1.5. The last column reports the slopes of first overtone PL relations over the whole period range covered by models.}
\label{tabPL}
\begin{tabular}{lcccc|c}
\hline
\hline
Band&b$_{F,all}$&b$_{F,short}$&b$_{F,long}$&b$_{F, \le 1.5}$&b$_{FO,all}$\\
$$Z=0.004$$\\
u&-2.503(0.016)&-3.134(0.016)&-0.897(0.088)&-2.752(0.014)&-2.963(0.015)\\
g&-2.893(0.012)&-3.362(0.015)&-1.789(0.066)&-3.065(0.012)&-3.136(0.016)\\
r&-3.098(0.009)&-3.440(0.011)&-2.301(0.050)&-3.222(0.009)&-3.291(0.012)\\
i&-3.177(0.008)&-3.469(0.010)&-2.494(0.043)&-3.281(0.008)&-3.363(0.010)\\
z&-3.231(0.007)&-3.494(0.009)&-2.620(0.038)&-3.326(0.007)&-3.410(0.009)\\
\hline
$$Z=0.008$$\\
u&-2.274(0.016)&-2.875(0.016)&-0.849(0.088)&-2.498(0.015)&-2.751(0.013)\\
g&-2.707(0.012)&-3.131(0.014)&-1.766(0.065)&-2.855(0.012)&-2.919(0.014)\\
r&-2.949(0.099)&-3.257(0.011)&-2.265(0.049)&-3.056(0.009)&-3.080(0.010)\\
i&-3.044(0.008)&-3.309(0.009)&-2.459(0.042)&-3.137(0.008)&-3.155(0.008)\\
z&-3.109(0.007)&-3.346(0.009)&-2.588(0.038)&-3.192(0.007)&-3.206(0.007)\\
\hline
$$Z=0.02$$\\
u&-1.156(0.0232)&-1.833(0.038)&-0.361(0.070)&-1.503(0.025)&-2.908(0.024)\\
g&-1.973(0.015)&-2.307(0.031)&-1.558(0.047)&-2.151(0.018)&-3.101(0.024)\\
r&-2.396(0.011)&-2.632(0.023)&-2.092(0.035)&-2.527(0.013)&-3.299(0.018)\\
i&-2.562(0.10)&-2.763(0.020)&-2.306(0.030)&-2.672(0.010)&-3.390(0.016)\\
z&-2.680(0.009)&-2.852(0.017)&-2.460(0.027)&-2.774(0.007)&-3.449(0.014)\\
\hline

\hline
\end{tabular}								
\end{center}				
\end{table*}

\subsection{The Wesenheit relations}
As the PL relation  reflects the instrinsic width of the instability strip and
its application implies the availability of a statistically significant
sample, as well as the need to disentangle reddening and
metallicity effects on distance determinations \citep[see
discussions in][]{cmr99,cmms00,cmmp01}, the reddening free Wesenheit
relation is often preferred. 
The latter relation relies on an assumed reddening law and includes a color term
similar, but not identical, to the Period-Luminosity-Color color term (see below), thus
partially reducing the effect of the finite width of the instability strip \citep[see][]{cmm00}.
The predicted Wesenheit functions in the SDSS filters have been built
by adopting the color coefficients given by \citet{girardi2004}
and are reported in Table 3  for both the fundamental and the first
overtone mode.
These results clearly indicate the decrease in the intrinsic
dispersion when passing from the PL to the Wesenheit relations, in
agreement with previous empirical and theoretical results \citep[][]{mf1991,cmm00,r12}. 
The obtained Wesenheit relations for Z=0.004 are shown in Fig. 5. These plots confirm the linearity and the small intrinsic dispersion of the
Wesenheit relations that make them powerful tools to infer Galactic
and extragalactic Cepheid distances.
\begin{figure}
\includegraphics[width=9cm]{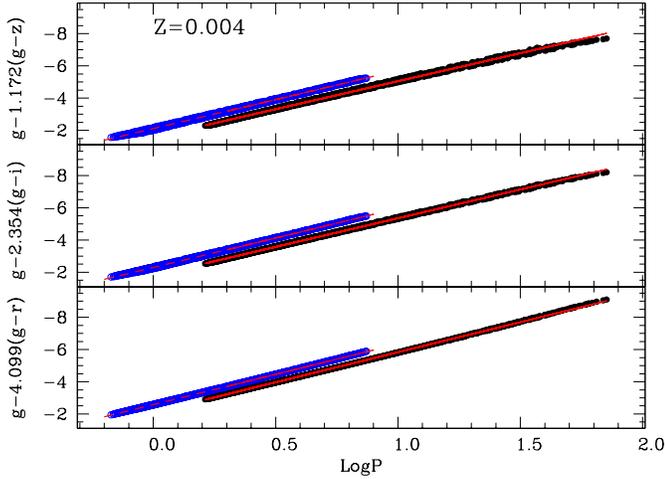} 
\vspace{-2cm}
\caption{Wesenheit relations in different SDSS photometrical bands for fundamental (solid line) and first overtone (dashed line) models with Z=0.004 The same figures for  Z=0.008 and Z=0.02 are given in the  on-line version of the paper.}
\label{figW1}
\end{figure}
\begin{table}
\begin{center}
\caption{ The Wesenheit relations  for fundamental and first overtone pulsators with different metallicities.}
\begin{tabular}{lcccc}
\hline
\hline
$$W=a +b logP$$\\
W&&a&b&rms\\
\hline
\hline
$$Z=0.004$$\\ 
g-4.099(g-r)      &F &    -2.01   &     -3.73   &  0.02 \\
                  &FO&    -2.58   &     -3.77     & 0.01 \\
                                                 
g-2.354(g-i)     &F &    -1.79    &   -3.56     &  0.03  \\
                 &FO&     -2.28   &   -3.67      &  0.03 \\

g-1.722(g-z) &F &  -1.58     & -3.48    &       0.06\\ 
            &FO&    -2.10   & -3.61     &      0.04\\

\hline

$$Z=0.008$$ \\
g-4.099(g-r)&F &    -2.11    &     -3.70 &      0.02\\
            &FO&    -2.55    &     -3.58 &     0.01\\
                                     
g-2.354(g-i)&F &     -1.79   &     -3.50 &       0.04\\ 
            &FO&     -2.25   &     -3.47 &      0.01\\

g-1.722(g-z) &F & -1.58       &  -3.40    &       0.06\\ 
            &FO&  -2.07       &  -3.41    0.03\\

\hline                 

$$Z=0.02$$\\
g-4.099(g-r)&F &     -2.131  &    -3.71   &    0.015\\
            &FO&     -2.641  &    -3.91   &   0.007\\
                                              
g-2.354(g-i)&F &     -1.89   &    -3.36   &    0.03\\
            &FO&     -2.32   &    -3.78   &    0.02\\

g-1.722(g-z) &F & -1.70      & -3.19     &       0.05\\ 
            &FO&   -2.11    & -3.70     &      0.03\\

\hline

\end{tabular}								
\end{center}				
\end{table}

We notice that in agreement with previous results in the
Johnson-Cousins bands \citep[][]{cmm00}, the metallicity dependence of
the Wesenheit relations is reversed when passing from optical to
near-infrared colors with the smallest effect for the $g$, $g-r$ combination.

\subsection{The Period-Luminosity-Color relations}

The Wesenheit relations are reddening free but only partially correcting the effect of the
finite width of the instability strip with the inclusion of their
color term. However, the true relation that holds for each individual
Cepheid is the Period-Color-Luminosity (PLC) relation, directly
descending from the combination of the period-density, the
Stephan-Boltzmann and the evolutionary Mass-Luminosity relations
\citep[see e.g.][and references therein]{bccm99,m09}.
The disadvantage is that  both individual reddening and multiband photometry are required in order to apply PLC relations to infer Cepheid distances and this is the reason why these relations are rarely adopted in the literature.
In any case the theoretical SDSS PLC relations based on the adopted
pulsation and atmosphere models are reported in Table 4 and shown in
Figs. 6, 7 and 8. We note that, differently from the PL and the Wesenheit
relations, the PLC ones have been derived from the individual pulsating
models and not from the simulated Cepheid distributions within the
respective instability strips. This because of the mentioned intrinsic nature of
these relations, expected to hold for each individual pulsator.
 Again we find a smaller metallicity effect in the case of first
 overtone models.

\begin{figure}
 \includegraphics[width=9cm]{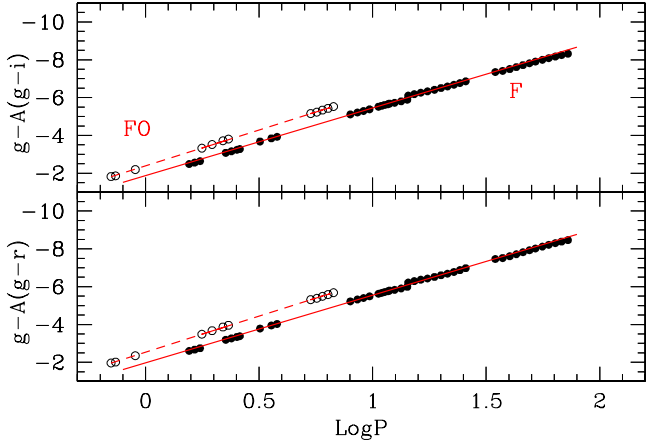} 
\vspace{-3cm}
\caption{PLC relations for F (closed circles) and FO (open circles) models with Z=0.004. See also Table \ref{tablePLC}.}
\label{fig1PLC}
\end{figure}
\begin{figure}
\includegraphics[width=9cm]{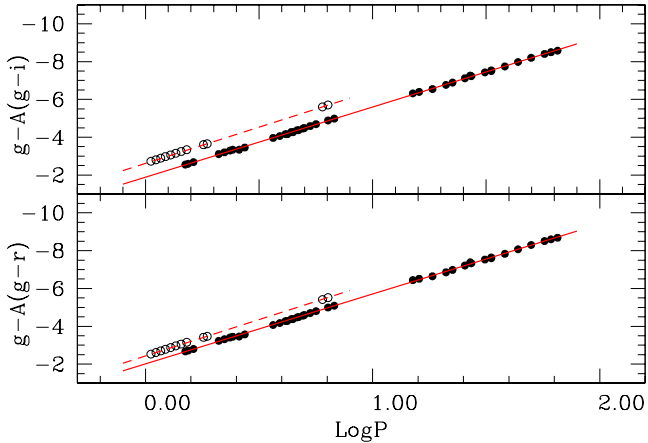}
\vspace{-3cm}
\caption{As for Fig. \ref{fig1PLC} but for Z=0.004}
\label{fig2PLC}
\end{figure} 
\begin{figure}
\includegraphics[width=9cm]{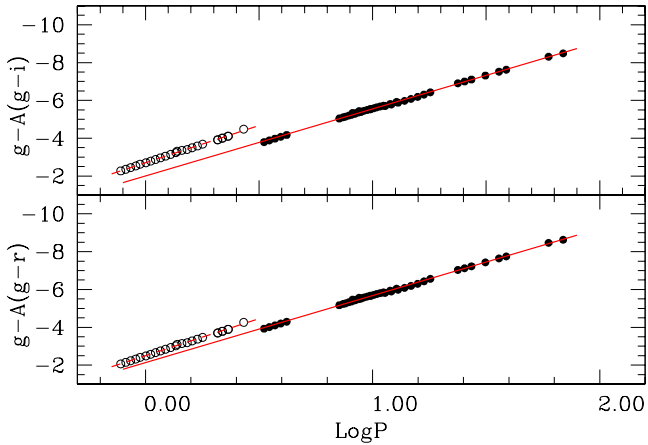}
\vspace{-3cm}
\caption{As for Fig. \ref{fig1PLC} but for Z=0.02}
\label{fig3PLC}
\end{figure}

\begin{table}
\begin{center}
\caption{PLC Relations for fundamental and first overtone pulsators with different metallicities.}
\label{tablePLC}
\begin{tabular}{lc|cccc}
\hline
\hline
$$g=a $\it colour$ +b logP+c$$\\
&$\it colour$ &a&b&c&rms\\
\hline
\hline
$$Z=0.004$$\\ 
F &(g-r)&  3.99&-3.65 & -2.06  &0.04 \\
  &(g-i)&  2.87&-3.66 &  1.96   &0.04 \\
FO &(g-r)& 4.01&-3.80 & -2.54  &0.02 \\
   &(g-i)& 2.79&-3.79 & -2.39  &0.02 \\
\hline
$$Z=0.008$$\\ 
F  &(g-r)& 3.99 &-3.70 &-2.01&0.03 \\
   &(g-i)& 2.85 &-3.71 &-1.89&0.03  \\
FO &(g-r)& 4.32 &-3.83 &-2.62 &0.01\\
   &(g-i)& 2.96 &-3.83 &-2.44 & 0.01\\
\hline                 
$$Z=0.02$$\\
F &(g-r)& 3.86 &-3.55 &-2.13  &0.05 \\
  &(g-i)& 2.75 &-3.55 &-2.00  &0.04 \\
FO&(g-r)& 4.32 &-3.93 &-2.71  &0.03\\
  &(g-i)& 2.93 &-3.92 &-2.50  &0.03\\
\hline
\end{tabular}
\end{center}
\end{table}

\section{Comparison with previous relations in the literature}\label{comp}

In a recent paper, \citet[][]{nk07} have derived semiempirical Cepheid
PL relations in the SDSS magnitudes by combining observed LMC BVI
mean magnitudes with theoretical bolometric corrections.  In the same
paper these authors construct empirical $g$,$r$ PL relations, using the
publicly available Johnson-Sloan photometric transformations, finding
a good agreement with the semiempirical ones. 

In Fig. 9 we show the comparison between the PL slopes derived in
this paper for fundamental models with $Z=0.008$ and $\log P < 2.0$
(filled symbols) and the semiempirical ones
\citep[][open symbols]{nk07} for the $u,g,r,i,z$ filters .
In the case of $g$ and $r$ the empirical relations by the same authors
(asterisks) are also shown.
These comparisons suggest that the slope of the PL relations obtained
in this paper, using suitable pulsation and atmosphere models,  for periods shorter than 100 d, are systematically
steeper than the semiempirical and empirical ones presented by
\citet{nk07}, with a deviation decreasing from $\sim$ 13 $\%$ in the $u$
filter to $\sim 3 \%$ in the $z$ filter.

\section{Conclusions}

In this paper we have provided the first theoretical scenario to our
knowledge for the interpretation of Classical Cepheid properties
observed in the SDSS filters. Extensive and detailed sets of nonlinear
convective pulsation models computed for the typical chemical
compositions of Galactic and Magellanic pulsators have been
transformed into the SDSS filters by using bolometric corrections
and color-temperature transformations based on updated model
atmospheres.
The main results of this investigation can be summarized as follows:
\begin{itemize}

\item The transformed instability strip confirms the nonlinearity of
the fundamental boundaries and the dependence on the metal content
already found in the Johnson-Cousins bands. This behaviour explains why  PL relations are not linear at large periods. 

\item The PL relations in the SDSS filters show evidence of the break at 10 days, in particular for the
lower metal contents and the shorter wavelengths, as expected on the
basis of previous empirical and theoretical studies. 

\item The theoretical light curves in the SDSS filters present morphological
characteristics similar to the ones in the Johnson-Cousins bands, with
evidence of the Hertzprung-progression phenomenon at the expected
period for each selected chemical composition.

\item The derived Wesenheit and PLC relations have, as expected, a smaller
  intrinsic dispersion than PL relations and show a metallicity dependence
  that resembles the behaviour of their Johnson-Cousins counterparts.

\item The obtained relations for first overtone pulsators are all
  linear and  much
  less affected by metallicity than the fundamental mode ones.

\item A comparison between the relations obtained in this paper and
  the semiempirical and empirical ones obtained by \citet{nk07}
  indicates a discrepancy in the slopes ranging from  $\sim$ 13 $\%$ in  $u$
 to $\sim 3 \%$ in $z$, with the theoretical relations
being systematically steeper.

\end{itemize}

\begin{figure}
\includegraphics[width=9cm]{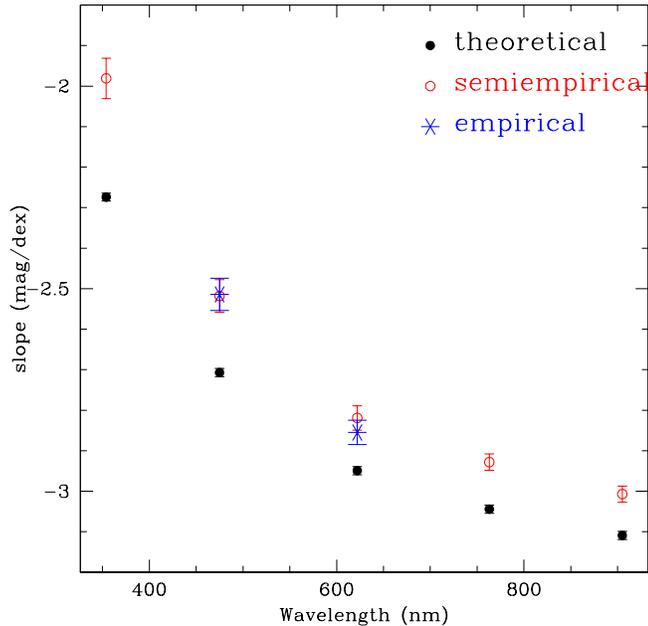} 
\caption{Comparison between the slopes of the theoretical PL relations
derived in this paper and the semiempirical and empirical ones by \citet{nk07}.}
\label{fig13}
\end{figure}

{}

\begin{thebibliography}{}

\bibitem[Abazajian et al.(2003)]{ab03} Abazajian, K., 
Adelman-McCarthy, J.~K., Ag{\"u}eros, M.~A., et al. 2003, AJ, 126, 2081 

\bibitem[Abbott et al.(2006)]{ab2006} Abbot, T., M., C., 2006, SPIE.6267E.119A

\bibitem[Alexander \& Ferguson(1994)]{af94} Alexander, D. R., Ferguson, J. W. 1994, ApJ, 437, 879

\bibitem[Bono et al.(1998)]{b98} Bono, G., Caputo, F., 
\& Marconi, M.\ 1998, ApJL, 497, L43 

\bibitem[Bono et al.(1999a)]{bms99} Bono, G., Marconi, M.,  \& Stellingwerf, R.~F.\ 1999, ApJSS, 122, 167 

\bibitem[Bono et al.(1999b)]{bccm99} Bono, G., Caputo, F., Castellani, V., \& Marconi, M.\ 1999, ApJ, 512, 711 

\bibitem[Bono et al.(2000a)]{bms00} Bono, G., Marconi, M., \& Stellingwerf, R.~F.\ 2000, A\&A, 360, 245 

\bibitem[Bono et al.(2000b)]{bcm00} Bono, G., Castellani, V., 
\& Marconi, M.\ 2000, ApJ, 529, 293 


\bibitem[Bono et al.(2001)]{b01} Bono, G., Gieren, W.~P., 
Marconi, M., Fouqu{\'e}, P., \& Caputo, F.\ 2001, ApJ, 563, 319 

\bibitem[Bono et al.(2002)]{b02} Bono, G., Groenewegen, 
M.~A.~T., Marconi, M., \& Caputo, F.\ 2002, ApJL, 574, L33 

\bibitem[Bono et al.(2000)]{bms00} Bono, G., Marconi, M., \& Stellingwerf, R.~F.\ 2000, A\&A, 360, 245 


\bibitem[Bono(2010)]{bo10} Bono, G. 2010, MmSAI, 81, 863


\bibitem[Bono \& Stellingwerf(1994)]{bs94} Bono, G., Stellingwerf, R.F. 1994, ApJS, 93, 233                   

\bibitem[Caputo et al.(2000)]{cmm00} Caputo, F., Marconi, M., \& Musella, I.\ 2000, A\&A,
354, 610 

\bibitem[Caputo et al.(2001)]{cmmp01} Caputo, F., Marconi, M., Musella, I., \& Pont, F. 2001, A\&A, 372, 544 

\bibitem[Caputo et al.(2000)]{cmms00} Caputo, F., Marconi, M., Musella, I., \& Santolamazza, P.\ 2000, A\&A, 359, 1059 

\bibitem[Caputo et al.(1999)]{cmr99} Caputo, F., Marconi, M., \& Ripepi, V.\ 1999, ApJ, 525, 784 

\bibitem[Castelli, Gratton \& Kurucz(1997a)]{cast97a} Castelli, F., Gratton, R. G., \& Kurucz, R. L. 1997a, A\&A, 318, 841 

\bibitem[Castelli, Gratton \& Kurucz(1997b)]{cast97b} Castelli, F., Gratton, R. G., \& Kurucz, R. L. 1997b, A\&A, 324, 432 

\bibitem[Chiosi, Wood \& Capitano(1993)]{chiosi1993}Chiosi, C.,  Wood, P. R. \& Capitanio, N., 1993, ApJS, 86, 541C

\bibitem[Di Criscienzo, Marconi \& Caputo (2004)]{dmc04} Di Criscienzo, M., Marconi, M., Caputo, F. 2004, ApJ, 612, 1092

\bibitem[Freedman et al.(2001)]{f01} Freedman, W.~L., Madore, B.~F., Gibson, B.~K., et al.\ 2001, ApJL, 553, 47 

\bibitem[Giradi et al.(2004)]{girardi2004}Girardi, L.,Grebel, E. K., Odenkirchen, M., Chiosi, C., 2004,A\&A, 422, 205G

\bibitem[Hertzsprung(1926)]{H26} Hertzsprung, E. 1926, Bull. Astron. Inst. Netherlands, 3, 115

\bibitem[Ledoux \& Walraven(1958)]{LW58} Ledoux, P., \& Walraven, T.\ 1958, Handbuch der Physik, 51, 353 

\bibitem[Kaiser(2993)]{K2003} Kaiser, N., 2004, SPIE, 5489, 11K

\bibitem[Madore(1982)]{m82} Madore, B.~F.\ 1982, ApJSS, 253, 575 

\bibitem[Madore \& Freedman(1991)]{mf1991} Madore, B. F., \& Freedman, W. L. 1991, PASP, 103, 933

\bibitem[Marconi et al.(2005)]{mmf05} Marconi, M., Musella,  I., \& Fiorentino, G.\ 2005, ApJ, 632, 590 

\bibitem[Marconi(2009)]{m09} Marconi, M.\ 2009,  MSAIT,80, 141 

\bibitem[Ngeow 
\& Kanbur(2007)]{nk07} Ngeow, C., \& Kanbur, S.~M.\ 2007, ApJ, 667, 35 


\bibitem[Ngeow et al.(2008)]{n08} Ngeow, C., Kanbur, S.~M., \& Nanthakumar, A.\ 2008, A\&A, 477, 621 
 \
\bibitem[Ripepi et al.(2006)]{ripepiSTEP} Ripepi, V.,    Marconi, M., Musella, I. et al. 2006, MSAIS, 9, 267

\bibitem[Ripepi et al.(2012)]{r12} Ripepi, V., Moretti, 
M.~I., Marconi, M., et al.\ 2012, MNRAS, 3283 

\bibitem[Saha et al.(2001)]{s01} Saha, A., Sandage, A., 
Tammann, G.~A., et al.\ 2001, ApJ, 562, 314

\bibitem[Sandage et  al.(2004)]{s04} Sandage, A., Tammann, G.~A., \& Reindl, B.\ 2004, A\&A, 424, 43  

\bibitem[Stellingwerf(1982)]{s82} Stellingwerf, R. F. 1982, ApJ, 262, 330

\bibitem[Tyson(2002)]{t2002} Tyson, J., A.,  2002, SPIE.4836, 10T



\end{thebibliography}
\end{document}